\documentclass[12pt, a4paper]{article}

    \usepackage{cmap} % Better search and copypasting of cyrillic text from pdf-file. Can cause errors if used together with pdfx package.
\usepackage[T1]{fontenc}
\usepackage[utf8]{inputenc}
\usepackage[english]{babel}

\usepackage[margin=2cm]{geometry}

\usepackage{microtype}

\usepackage{amsmath, amssymb, amsthm}
\usepackage{indentfirst}
\usepackage[sort,compress]{cite}
\usepackage[affil-it]{authblk}

\usepackage[margin=0.2em]{subcaption}

\usepackage{graphicx}

\usepackage{hyperref}

\graphicspath{{figures/}}

\DeclareMathOperator*{\argmin}{argmin}
\DeclareMathOperator{\tr}{tr}
\DeclareMathOperator{\bydef}{\overset{\text{def}}{=}}
\DeclareMathOperator{\Expect}{\mathbb{E}}

\theoremstyle{remark}
\theoremstyle{definition}

\newtheorem{example*}{Пример}

\newtheorem*{remark*}{Замечание}
\newlength{\subfigsize}
\newcommand{\vq}{\mathbf{q}}
\newcommand{\vrho}{\boldsymbol{\rho}}
\newcommand{\vr}{\mathbf{r}}

\newcommand{\prior}{\textnormal{pr}}

\title{Object reconstruction from multiplexed quantum ghost images
using reduction technique}
\author[1]{D.\,A.\,Balakin\thanks{Corresponding author e-mail: balakin\_d\_a@physics.msu.ru}}
\author[1]{A.\,V.\,Belinsky}
\author[1,2]{A.\,S.\,Chirkin}

\affil[1]{M.\,V.\,Lomonosov Moscow State University, Faculty of Physics, Leninskie Gory, 1, bld 2, Moscow 119991, Russia}
\affil[2]{M.\,V.\,Lomonosov Moscow State University, The International Laser Center, Leninskie Gory, 1, bld 62, Moscow 119991, Russia}
\date{}
\begin{document}

\maketitle

\begin{abstract}
We apply the measurement reduction technique
to optimally reconstruct an object image
from multiplexed ghost images (GI)
while taking into account both GI correlations
and object image sparsity.
We show that one can reconstruct an image in that way
even if the object is illuminated by a small photon number.
We consider frequency GI multiplexing
using coupled parametric processes.
We revealed that the imaging condition depends on
the type of parametric process, namely, whether down- or up-conversion is used.
Influence of information about sparsity in
discrete cosine transform and Haar transform bases
on reconstruction quality is studied.
In addition, we compared ordinary and ghost images
when the detectors are additionally illuminated by noise photons
in a computer experiment, which showed increased noise immunity of GI,
especially with processing via the proposed technique.
\end{abstract}

\section*{Introduction}

By now, to enhance human visual capability
a vast high-tech base
including highly sensitive, high-precision and high-speed cameras
have been developed.
Nevertheless, there still are objects
whose direct optical observation is difficult.
They are primarily halftone biological objects
that are especially sensitive to illumination
and thus, have to be investigated very delicately.
Ghost imaging (GI) are one way of solving this problem,
as it allows to obtain object images
without direct observation of its spatial structure.
For GI, correlated light beams are necessary.
GI enables extraction of object information
from spatial correlations between beams,
one of which (in the object arm) interacts with the object,
while the other one (in the reference arm) does not.
In the object arm, a bucket detector is used,
which provides only information about the total intensity of the transmitted radiation.
The other beam does not interact with the object, but is detected by a CCD matrix,
which permits measuring the spatial correlation function of intensity between two arms.
The information about transparency or reflectivity distribution of the research object
is extracted from photocount correlations in the object and reference arms \cite{belinsky_klyshko_1994},
see also \cite{gatti_et_al_2004, gatti_et_al_2007, chan_et_al_2009, erkman_shapiro_2010, shapiro_boyd_2012}.

In this paper, we study application of
multicomponent entangled quantum light states
that let us produce several GI simultaneously
(to multiplex GI)~\cite{chirkin_jetp_letters_15, mgi_icono_lat_2016, mgi_correlations_jrlr_2017, ghost_images_jetp}
by using radiation with different frequencies in reference arms.
Mutual correlations of multiplexed images are used as additional information
to improve image processing in the presence of fluctuations.
There are various ways of producing multi-frequency entangled light beams.
The required states can be obtained, e.\,g.,
in consecutive coupled parametric interactions in nonlinear crystals
located either outside~\cite{rodionov_chirkin_2004, ferraro_et_al_2004}
or inside~\cite{olsen_drummond_2005} an optical resonator,
in nonlinear waveguide structures~\cite{solntsev_et_al_2012, kruse_et_al_2013}
where modes are coupled through evanescent modes,
in a spatially modulated pump beam~\cite{daems_et_al_2010}.
The considered GI multiplexing employs four-frequency entangled quantum states
formed through parametric decay of pump photons
into two photons with different frequencies
that are mixed in the same crystal with pump photons,
which produces photons with sum frequencies~%
\cite{chirkin_shutov_2007, chirkin_shutov_2009}.
Quantum theory of this process has been systematically developed in recent years
\cite{saygin_chirkin_2010, saygin_chirkin_2011, saygin_et_al_2012, tlyachev_et_al_2013}.
Note that in~\cite{duan_et_al_2013, zhang_et_al_2015, chan_et_al_2009}
GI were multiplexed via multi-frequency noncoherent radiation sources
to simultaneously produce several GI that are superimposed afterwards.
Recently, polarization multiplexing
has been used in several works on GI
(see \cite{polar_mult_gi} and references there),
in particular, to improve the reconstructed image quality.

The ghost image processing methods
considered in the literature
usually rely on regularization.
The regularizing functional is
a characteristic of image sparsity in a given basis~%
\cite{imaging_small_n_photons,compressive_ghost_imaging,gong_han_2012, hi-res_gi_sparsity},
and the minimized functional itself is the least squares one~%
\cite{compressive_ghost_imaging, gong_han_2012, hi-res_gi_sparsity}
or likelihood function~\cite{imaging_small_n_photons}.
Alternatively,
a sparsity characteristic (e.\,g. the $L^1$ norm in a given basis) is minimized~\cite{katz_et_al}
under the constraint that measuring the image reconstructed in that way
would give the results actually obtained.
Since such functional is not connected the error of the interpretation result,
the obtained estimate is, generally speaking, not the optimal one.
Unlike \cite{imaging_small_n_photons},
measurement reduction method, including its proposed version,
does not require only Poisson photocount distribution,
and unlike \cite{imaging_small_n_photons, compressive_ghost_imaging, katz_et_al, gong_han_2012, hi-res_gi_sparsity},
image sparsity in any basis is not required.

Note the main differences between this article and publications \cite{mgi_icono_lat_2016, mgi_correlations_jrlr_2017, ghost_images_jetp},
in which multiplexed GI processing using measurement reduction technique
was employed as well.
Firstly, in these works the situation was considered when
the only information about transparency distribution available to the researcher
was that its values belong to a unit interval.
In this article, it is considered that the researcher also has information about
transparency distribution sparsity in a given basis
and wants to take advantage of it to improve estimation quality.
Secondly, as this information enables reconstruction of acceptable quality
even with a small number of photons illuminating the object,
multiplexed ghost imaging with a small number of photons
($\sim 1 \div 10$~photons per pixel) and processing of acquired images
is modeled (see Sec.~\ref{sec:computer-modelling}).
Thirdly, the presented version of measurement reduction technique
differs from the one used in \cite{mgi_icono_lat_2016, mgi_correlations_jrlr_2017, ghost_images_jetp}
in that projection (to take into account the information about the object)
minimizes Mahalanobis distance instead of Euclidean distance,
see Eqn.~\eqref{eqn:mahalanobis-projection} below.
Finally, fourthly,
in the studied multiplexed ghost imaging setup
the object arm is lensless.
This leads to imaging conditions
depending on the type of parametric coupling of photon frequencies
in the object arm and the reference arms.

The article structure is as follows.
In section~\ref{sec:imaging}, we discuss GI multiplexing setup
with lensless object arm and with lenses in reference arms.
In section~\ref{sec:image-processing}
the measurement reduction method is outlined.
The information about the object that is available to the researcher
and that is employed in reduction is summarized in
subsection~\ref{sec:image-information}.
In subsection~\ref{sec:reduction-algorithm},
the algorithm of GI processing using reduction method
that takes this information into account
is described.
Computer modeling results are given
in section~\ref{sec:computer-modelling}.
Main results of the article are summarized in the conclusion.

\section{Frequency multiplexing of quantum ghost images}
\label{sec:imaging}

\begin{figure}
\centering
\includegraphics[width=\linewidth]{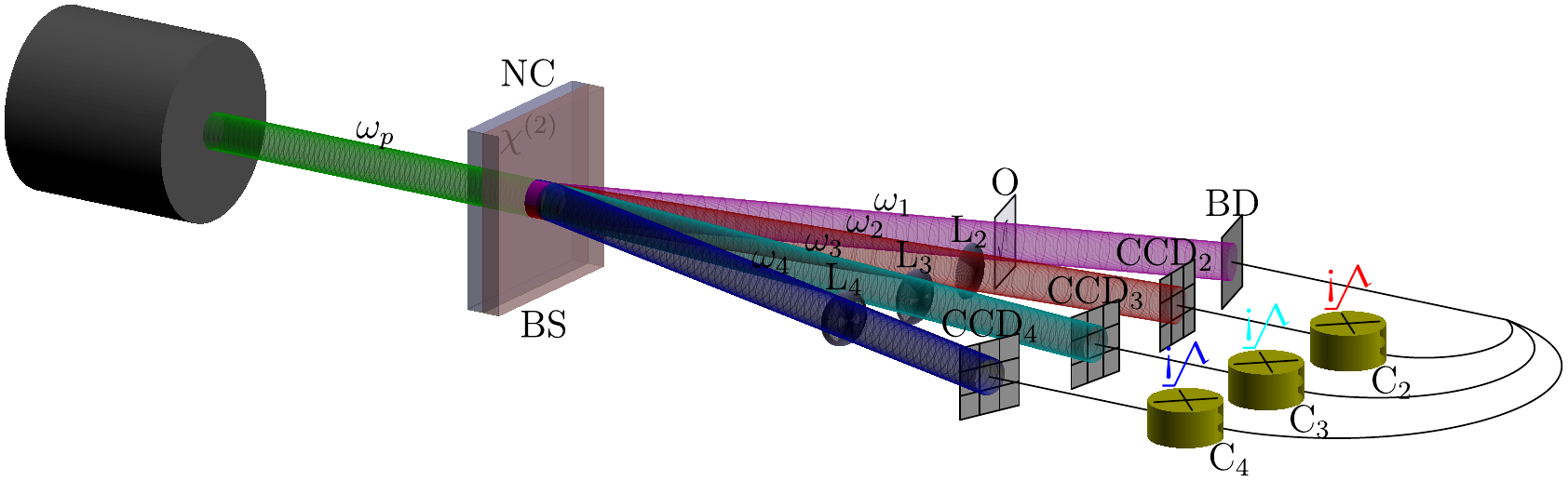}
\caption{Multiplexed ghost imaging setup.
NC is the nonlinear convertor;
BS is the beam splitter;
$\omega_{p}$ is pump frequency;
$\omega_{1}, \ldots, \omega_{4}$ are frequencies of produced entangled photons;
$O$ is the object;
BD is the bucket detector in the object arm;
$L_j$ are lenses with focal lengths $f_j$;
CCD$_j$ are CCD in reference arms;
C$_j$ are intensity correlators,
$j = 2, 3, 4$}
\label{fig:ghost-imaging}
\end{figure}

GI multiplexing setup is shown in fig.~\ref{fig:ghost-imaging}.
The illumination is provided by coupled parametric processes
that produce four-frequency entangled light fields.
Pump radiation incident into the nonlinear convertor (nonlinear photon crystal)
has frequency $\omega_{p}$.
In the crystal, pump photons decay into two photons
with related frequencies $\omega_1$ and $\omega_2$:
$\omega_p = \omega_1 + \omega_2$.

Four-frequency fields appear as a result of further conversion
of a part of photons with frequencies $\omega_{1}$ and $\omega_{2}$
to photons with frequencies $\omega_{3}$ and $\omega_{4}$
in frequency-mixing processes:
\begin{equation}
\label{eqn:mixing}
\begin{aligned}
\omega_{p} + \omega_{1} &= \omega_{3},\\
\omega_{p} + \omega_{2} &= \omega_{4}.
\end{aligned}
\end{equation}

Efficient energy exchange between interacting light waves in these processes
can be achieved in aperiodically nonlinear photon crystals,
e.\,g. in $\mathrm{Li Nb O}_3$,
in the in quasi-phase matched regime,
in which phase matching $\Delta k_j$ between interacting waves
are compensated by the vectors of the inverse nonlinear lattice
\cite{chirkin_shutov_2007, chirkin_shutov_2009}.
Note that the considered process was recently realized
via a setup with two nonlinear photon crystals
in the work \cite{suchowski_et_al_2016},
where the spectrum of a photon pair at frequency above pump frequency was studied.

Let ghost images be obtained by means of the optical system setup
shown in fig.~\ref{fig:ghost-imaging},
where a detector integrating radiation over the entire aperture
is used in the object arm.
It is assumed that the length of nonlinear photon crystal is chosen
so that the transversal wave number amplification band of the nonlinear convertor
substantially exceeds the width of wave spectrum of the object image.
Details of four-frequency entangled state formation were considered in~%
\cite{chirkin_jetp_letters_15, mgi_icono_lat_2016, mgi_correlations_jrlr_2017, ghost_images_jetp}.

In the setup of fig.~\ref{fig:ghost-imaging}
the object is illuminated by radiation with frequency $\omega_1$,
which is detected by the bucket detector (BD) over the entire beam aperture,
and therefore lacks spatial resolution.
Radiation with other frequencies $\omega_{2}, \omega_{3}, \omega_{4}$
after their spatial separation
enters reference arms with lenses in them.
Focal lengths $f_j$ of lenses
and their positions between the beam splitter (BS) and CCD cameras
are chosen according to imaging conditions.
These conditions depend on
the type of relation to the frequency of object illumination.
For frequencies $\omega_{2}$, $\omega_{4}$ the imaging condition has the form
(cf. the analogous condition in \cite{belinsky_klyshko_1994}
for equal frequencies)
\begin{equation}
\label{eqn:image24}
\frac{1}{f_j} = \frac{1}{l_{j2}} + \frac{1}{l_{j1} + (\lambda_{1}/\lambda_{j})l_{11}},
j = 2, 4,
\end{equation}
while for frequency $\omega_3$ it reads
(cf. the analogous condition in \cite{vyunishev_et_al_2015}
for equal frequencies)
\begin{equation}
\label{eqn:image3}
\frac{1}{f_3}= \frac{1}{l_{32}}+\frac{1}{l_{31}-(\lambda_{1}/\lambda_{3})l_{11}}.
\end{equation}
In expressions \eqref{eqn:image24}, \eqref{eqn:image3}
the length $l_{11}$ is the distance from BS to the object,
$l_{j1}$ is the distance from BS to the lens,
$l_{j2}$ is the distance from the lens to the detector CCD$_j$,
$\lambda_j$ is the length of the wave of corresponding frequency.
The derivation of Eqns. \eqref{eqn:image24}, \eqref{eqn:image3}
is described below.
They are a generalization of known ones to the case of
different frequencies of radiation illuminating the object
and radiation in reference arms.

The theory of formation of entangled quantum four-beam states in processes \eqref{eqn:mixing}
was developed in \cite{saygin_chirkin_2010, saygin_chirkin_2011, saygin_et_al_2012}.
Fourier components of Bose operators of field at nonlinear crystal output
are represented in matrix form:
\begin{equation}
\label{eqn:8}
\hat{\mathbf{a}}(\vq, l) = Q(\vq, l){\hat{\mathbf{v}}}(\vq),
\end{equation}

Here $\hat{\mathbf{a}}$ and $\hat{\mathbf{v}}$ are columns of Bose operators
at crystal output and input, respectively.
They are of the form
$\hat{\mathbf{a}} \bydef (\hat{a}_{1}, \hat{a}_{2}^{\dagger}, \hat{a}_{3}, \hat{a}_{4}^{\dagger})^{T}$,
where $T$ denotes transposition,
$\hat{a}_1 = \hat{a}_1(\vq, l)$,
$\hat{a}_{2}^{\dagger}= \hat{a}_{2}^{\dagger}(-\vq, l)$,
$\hat{a}_3 = \hat{a}_3(\vq, l)$,
$\hat{a}_{4}^{\dagger} = \hat{a}_{4}^{\dagger}(-\vq, l)$,
$l$ is the length of nonlinear crystal.
Operators in the column
$\hat{\mathbf{v}} \bydef (\hat{v}_{1}, \hat{v}_{2}^{\dagger}, \hat{v}_{3}, \hat{v}_{4}^{\dagger})^{T}$
refer to vacuum field state.

$Q$ is a $4 \times 4$ matrix whose elements $Q_{mn}$ describe field conversion
from frequency $\omega_n$ to frequency $\omega_m$.
The form of the matrix $Q$ and its properties in the quasioptical approximation
are given in \cite{saygin_chirkin_2010}.
The elements of $Q$ depend on crystal length, pump intensity and transversal wave number $\vq$.

The opertor $\hat{a}_{j}(\vq, z)$
is the annihilation operator of
plane mode photons with frequency $\omega_j$ and transversal wave vector $\vq$:
\begin{equation}
\label{eqn:7}
\hat{a}_{j}(\vq, z) = \frac{1}{2\pi} \iint\limits_{-\infty}^{+\infty} \hat{A}_j (\vrho, z) \exp(-i \vq \vrho) d\vrho,
\end{equation}
where $\hat{A}_j (\vrho, z)$ is the slowly varying amplitude operator of
the positive-frequency field
\begin{equation}
\label{eqn:7a}
\hat{E}^\dagger_{j}(\vr, z, t) =  \hat{A}_j (\vr, z, t) \exp(-i(\omega_{j} t - k_{j}z)),
\end{equation}
$k_{j}$ is wave number,
$z$ is the direction of propagation of interacting waves in the nonlinear crystal.

After BS, their amplitude operators are defined by the following relations
\begin{equation}
\label{eqn:11}
\hat{B}_{j}(\vr_{j}) = \int H_{j}(\vr_j, \vrho_{j}){\hat{A}}_{j}(\vrho_{j}, l) d\vrho_{j},
\end{equation}
in the detector plane,
integration being over the light beam aperture.
$H_{j}(\vr_j, \vrho)$ is the medium response function
for radiation propagation from the crystal to the detector in $j$-th arm.
We assume for simplicity
that beam splitting takes place
directly at nonlinear crystal output.
In other words, BS is considered to be thin.

For the object arm
\begin{equation}
\label{eqn:resp1}
H_{1}(\vr_1, \vrho_{1}) = \int\limits_{-\infty}^{+\infty} H_{1}(\vr_1 - \vrho_{1}'; l_{12}) T(\vrho_{1}') H_{j}( \vrho_{1}' - \vrho_{1}; l_{11}) d\vrho_{1}'.
\end{equation}
Here $T(\vrho_{1}')$ is the object transmission coefficient,
$H_{1}(\vr - \vrho; l_{1j})$ is the Green's function
\begin{equation}
\label{eqn:resp2}
\int H_{1}(\vr - \vrho; l_{1j}) = -i\frac{k_1}{2\pi l_{1j}} \exp\left(i\frac{k_{1}(\vr - \vrho)^2}{2 l_{1j}}\right).
\end{equation}
As noted above, $l_{11}$ is the distance between BS and the object
and $l_{12}$ is the distance between the object and the bucket detector.

Response functions of reference arms containing thin lenses with focal length $f_j$
can be represented as (see \cite{intro_stat_radiophys_optics, goodman_fourier_optics})
\begin{equation}
\label{eqn:resp3}
H_{j}(\vr_j, \vrho_{j}) = -i \frac{k_j}{2\pi L_{j}} \exp\left(i\frac{k_{j}}{2 L_{j}}
\left[(\vr_j - \vrho_j)^2 - (l_{j1} \vr_j^2 + l_{j2} \vrho_j^2)/ f_j\right]
\right),
\end{equation}
where
\begin{equation*}
L_j = l_{j1} + l_{j2} - l_{j1} l_{j2}/f_j.
\end{equation*}

Intensity operators of the obtained beams are
$\hat{I}_j(\vr_j) = \hat{B}^\dagger_j(\vr_j) \hat{B}_j(\vr_j)$.
Mutual intensity correlation functions of the object arm and reference arms,
taking into account Gaussian field statistics,
are determined by the following formulas:
for radiation with frequency $\omega_3$
the correlation function is
\begin{equation}
G_{13}(\vr_1, \vr_3) =
\langle \hat{I}_1(\vr_1) \hat{I}_3(\vr_3) \rangle - \langle \hat{I}_1(\vr_1) \rangle \langle \hat{I}_3(\vr_3) \rangle =
| \langle \hat{B}_1 (\vr_1) \hat{B}_3^\dagger (\vr_3) \rangle |^2,
\end{equation}
while for radiation with frequencies $\omega_2$ or $\omega_4$
the correlation function is
\begin{equation}
G_{1j}(\vr_1, \vr_j) =
\langle \hat{I}_1(\vr_1) \hat{I}_j(\vr_j) \rangle - \langle \hat{I}_1(\vr_1) \rangle \langle \hat{I}_j(\vr_j) \rangle =
| \langle \hat{B}_1 (\vr_1) \hat{B}_j (\vr_j) \rangle |^2,
j = 2, 4.
\end{equation}
The difference in the definitions of the correlation functions under consideration
is due to the type of parametric conversion and the initial vacuum fluctuations.
As a consequence, only the vacuum operators in antinormal ordering
contribute to correlations.

Under imaging conditions \eqref{eqn:image24}, \eqref{eqn:image3}
the expressions can be transformed to
\begin{equation}
\label{eqn:corr-func-imag-cond-3}
G_{13}(\vr_1, \vr_3) =
| \Gamma_3 |^2 \left| \frac{l_{31} (\lambda_1/\lambda_3) l_{11}}{\lambda_1 l_{12} l_{32}}
T\left( - \alpha_3 \vr_3 \right) \right|^2,
\alpha_3 \bydef \frac{l_{31} - (\lambda_1/\lambda_3) l_{11}}{l_{32}},
\end{equation}
\begin{equation}
\label{eqn:corr-func-imag-cond-24}
G_{1j}(\vr_1, \vr_j) =
| \Gamma_j |^2 \left| \frac{l_{j1} (\lambda_1/\lambda_j) l_{11}}{\lambda_1 l_{12} l_{j2}}
T\left( - \alpha_j \vr_j \right) \right|^2,
\alpha_j \bydef \frac{l_{j1} + (\lambda_1/\lambda_j) l_{11}}{l_{j2}}.
\end{equation}
These formulas are derived under the assumption that at the nonlinear crystal output,
mutual correlation functions of the radiation
\begin{equation*}
\Gamma_{1j}(\vrho_1 - \vrho_j) =
\langle \hat{A}_1(\vrho_1, l) \hat{A}_j(\vrho_j, l) \rangle =
(2 \pi)^{-1} \int Q_{11j}(\vq) \exp( i \vq (\vrho_1 - \vrho_j)) d\vq,
j = 2, 4,
\end{equation*}
\begin{equation*}
\Gamma_{13}(\vrho_1 - \vrho_3) =
\langle \hat{A}_1(\vrho_1, l) \hat{A}_j^\dagger(\vrho_3, l) \rangle =
(2 \pi)^{-1} \int Q_{113}(\vq) \exp( -i \vq (\vrho_1 - \vrho_3)) d\vq,
\end{equation*}
where
$Q_{11n}(\vq) \bydef Q_{11}(\vq) Q_{n1}^{*}(\vq) + Q_{13}(\vq)Q_{n3}^{*}(\vq)$,
$n = 1, 2, 3$,
are substituted by $\delta$-functions
\begin{equation*}
\Gamma_{1n}(\vrho_1 - \vrho_n) = \Gamma_n \delta(\vrho_1 - \vrho_n),
\Gamma_n = \int \Gamma_{1n}(\vrho) d\vrho.
\end{equation*}
These substitutions are valid if the radiation correlation radius
is much smaller than a characteristic spatial scale of object image change.

The expressions \eqref{eqn:corr-func-imag-cond-3}, \eqref{eqn:corr-func-imag-cond-24} coincide, up to a factor before the image transmission coefficient,
with the expression obtained for another experimental setup \cite{mgi_correlations_jrlr_2017, ghost_images_jetp}.
In \cite{mgi_correlations_jrlr_2017, ghost_images_jetp}
ghost image correlations determined by fourth-order intensity correlations
(eight-order field ones)
were studied as well.
Obviously, in the setup under consideration they will be the same as in \cite{mgi_correlations_jrlr_2017, ghost_images_jetp}.
After integration over the area $s$ of the beam in the object arm (over $d\vr_1$)
intensity correlation functions of the second order, in accordance with \eqref{eqn:corr-func-imag-cond-3}, \eqref{eqn:corr-func-imag-cond-24}, becomes
\begin{equation}
\label{eqn:corr-func-2-order}
G_{j}(\vr_{j}) \sim
s
\left| T( - \alpha_j \vr_{j}) \right|^{2},
\end{equation}
while the GI correlation function determined by eighth-order field correlation function
becomes
\begin{equation}
\label{eqn:corr-func-4-order}
K_{ij}^{\textnormal{GI}}(\vr_{i}, \vr_{j}) \sim
s^2
|T(-\alpha_i \vr_i)|^2
|T(-\alpha_j \vr_j)|^2.
\end{equation}
Naturally, the coefficients $\alpha_2, \alpha_3, \alpha_4$
can be made equal by choice of setup parameters.
In addition, in the following formulas \eqref{eqn:a-op-form} and \eqref{eqn:sigma-op-form}
the factors dependent on the measurement unit choice
will be omitted for brevity.

As mentioned above, the correlation functions derived above
provide the information about the measuring (image acquisition) process
that is used in the measurement reduction technique
along with the information about the object.
The following section focuses on the measurement reduction technique itself
and the information about the object.
Not all of the information mentioned above is equal in importance:
only the correlation function \eqref{eqn:corr-func-2-order}
and finiteness of $L^2$ norm of the correlation function \eqref{eqn:corr-func-4-order}
for all $i, j = 2, 3, 4$
are strictly necessary for image reconstruction.
Nevertheless, additional information about both the measuring process
(in our case, the form of correlation function \eqref{eqn:corr-func-4-order})
and the object (sparsity of its transparency distribution)
can vastly improve reconstruction quality, as it will be shown below.

\section{Processing of acquired images}
\label{sec:image-processing}

The output of $i$-th correlator, denoted as $\xi^{(i)}(\vr)$,
can be considered as the impact of a measuring transducer (MT)
on the input signal $f(\vr) \sim |T(-\vr)|^2$.
Here and below, unlike the previous section,
$f$ denotes the vector describing the object transparency distribution
instead of focal length.
We assume for simplicity that $\alpha_2 = \alpha_3 = \alpha_4 = 1$.

We will consider piecewise constant images,
i.\,e. transparency is constant within each pixel.
Areas of constant transparency and constant brightness corresponding to pixels
are considered to be ordered in an arbitrary but fixed way.
Due to that it is sufficient for us to consider a finite number of values of $\vr$.
Thus, $f$ as the vector of transparencies is an element of
finite-dimensional Euclidean space $\mathcal{F}$.

An image processing algorithm
ought to provide the most accurate estimate of
the feature of the original image $f$ that is of interest to the researcher
based on obtained data $\xi$,
which consists of acquired ghost images $\xi^{(i)}(\vr)$, $i = 2, 3, 4$.
Measurement reduction method allows to obtain such an estimate.
Let us formulate the measurement model as
\begin{equation}
\label{eqn:measurement-model}
\xi = \mathbf{A} f + \nu,
\end{equation}
where $f$ is an priori unknown vector that describes the transparency distribution
of the object,
$\nu$ is measurement error with zero expectation, $\Expect \nu = 0$,
which means absence of systematic measurement error,
and covariance matrix
$\mathbf{\Sigma}_{\nu} = \Expect \nu \nu^*$.
The matrix $\mathbf{A}$ describes ghost imaging and GI acquisition:
the matrix element $\mathbf{A}_{ij}$ is equal to the mean output of $i$-th detector
for unit transparency of $j$-th element of the object and zero transparency of other object elements (i.\,e. whose indices differ from $j$).
The dimension of vector $f$ is the number of pixels in the object image,
while the dimension of $\xi$ is the number of pixels in all CCD.
The condition of systematic measurement error absense
$\Expect \nu = 0$
means, in particular, that the expectation of the component of measurement results
caused by detector dark noises is subtracted from the measurement results,
similar to \cite{gi_compressed_sensing_substr_const}.

Matrices $\mathbf{A}$ and $\mathbf{\Sigma}_{\nu}$ are related
to the correlation functions considered above.
The measuring setup employs correlators
that measure correlations between the object arm and other arms.
Therefore, the matrix $\mathbf{A}$, which models the impact of MT on the image,
is a block matrix and consists of three blocks
describing correlator outputs,
i.\,e. correlations between the object arm and reference arms:
\begin{equation}
\label{eqn:a-op-form}
\mathbf{A} = \begin{pmatrix}
\mathbf{B}_{2} \mathbf{C}_{2}\\
\mathbf{B}_{3} \mathbf{C}_{3}\\
\mathbf{B}_{4} \mathbf{C}_{4}
\end{pmatrix}.
\end{equation}
Under the conditions used to derive
the intensity correlation functions \eqref{eqn:corr-func-2-order} and \eqref{eqn:corr-func-4-order}
the matrices $\mathbf{C}_2$--$\mathbf{C}_4$ are identity ones multiplied by
pixel size and the factor before $|T(\vr_{i})|^{2}$
in expression~\eqref{eqn:corr-func-2-order} for the correlation function $G_{j}$.
The matrices $\mathbf{B}_2$--$\mathbf{B}_4$ model the detectors.
Specifically, the matrix element $(\mathbf{B}_i)_{pk}$ is equal to
the output of the detector in $i$-th arm at $p$-th position
for unit brightness of $k$-th pixel and zero brightness of other pixels.

Noise covariance matrix has block form as well:
\begin{equation}
\label{eqn:sigma-op-form}
\mathbf{\Sigma}_\nu =
\begin{pmatrix}
\mathbf{B}_2\mathbf{\Sigma}_{22}(f)\mathbf{B}_2^* & \mathbf{B}_2\mathbf{\Sigma}_{23}(f)\mathbf{B}_3^* & \mathbf{B}_2\mathbf{\Sigma}_{24}(f)\mathbf{B}_4^*\\
\mathbf{B}_3\mathbf{\Sigma}_{32}(f)\mathbf{B}_2^* & \mathbf{B}_3\mathbf{\Sigma}_{33}(f)\mathbf{B}_3^* & \mathbf{B}_3\mathbf{\Sigma}_{34}(f)\mathbf{B}_4^*\\
\mathbf{B}_4\mathbf{\Sigma}_{42}(f)\mathbf{B}_2^* & \mathbf{B}_4\mathbf{\Sigma}_{43}(f)\mathbf{B}_3^* & \mathbf{B}_4\mathbf{\Sigma}_{44}(f)\mathbf{B}_4^*
\end{pmatrix}
+ \mathbf{\Sigma}_{\nu'}.
\end{equation}
Here the element with indices $k$, $k'$
of the block~$\mathbf{\Sigma}_{ij}$ is equal to
the integral of $K_{ij}^{\textnormal{GI}}$ \eqref{eqn:corr-func-4-order}
over the values of~$\vr_{i}$ belonging to $k$-th pixel
and over the values of~\(\vr_{j}\) belonging to $k'$-th pixel,
for the same pixel ordering as in the matrix $\mathbf{A}$.
Hence, the dependence of \eqref{eqn:sigma-op-form} on $f$
is caused by the dependence on $|T(\cdot)|^2$ of the correlation function $K_{ij}^{\textnormal{GI}}$ \eqref{eqn:corr-func-4-order}.
The term $\mathbf{\Sigma}_{\nu'}$
is the covariance matrix of the noise component $\nu'$
that is unrelated to ghost imaging,
e.\,g. thermal noise in circuits and digitization error.
Most of the noise arising \emph{before} the correlators
is suppressed by them if noise in object and reference arms is independent,
but this does not apply to noise arising \emph{after} the correlators.
Besides, due to finite coincidence circuit match time
some of noise photons contribute to the noise as well,
see discussion of fig.~\ref{fig:gi-vs-ordinary-10} below.

It should be noted that the algorithm proposed below can be applied
for an image multiplexing method
that differs from the one considered in section~\ref{sec:imaging}
if the measurement model has the form~\eqref{eqn:measurement-model}.
Specifically the expectation of measurement result
has to be the product of a matrix $\mathbf{A}$
and the transparency distribution vector of the measured object,
and the error has to be able to be considered additive.
For that, the fourth-order intensity correlation function
(an analog of \eqref{eqn:corr-func-2-order})
has to linearly depend on the transparency distribution,
and the eighth-order intensity correlation function
(an analog of \eqref{eqn:corr-func-4-order})
has to ``sufficiently weakly'' depend on the transparency distribution
so that an unknown covariance matrix could be estimated using measurement results.
If, in addition to that,
photon detections in reference arms are conditionally independent
under fixed output of the bucket detector in the object arm
(output of a detector in a reference arm
does not affect output of detectors in other reference arms),
then what was said about the form of matrices $\mathbf{A}$ and $\mathbf{\Sigma}_{\nu}$
remains valid.

The estimation problem consists of reconstruction of the most accurate estimate
of the signal $\mathbf{U} f$ from the measurement result $\xi$,
where the matrix $\mathbf{U}$ describes
a measuring device that is ideal (for the researcher).
We consider the case when the researcher is interested in
reconstruction of the object image itself,
and imaging does not distort the object,
therefore, $\mathbf{U} = I$.

Since measurement results linearly depend on $f$,
to solve the estimation problem
we can use the model $[\mathbf{A}, \mathbf{\Sigma}_\nu, \mathbf{U}]$
described in \cite{pytyev_ivs}, see also \cite{pytyev_chulichkov_1998, pytyev_et_al_2004, pytyev_2010, reduction_vmu}.
If the estimation process is described by a linear operator $R$
($R \xi$ is the result of processing the measurement $\xi$),
the corresponding mean squared error (MSE)
in the worst case of $f$,
$h(R, \mathbf{U}) = \sup\limits_{f \in \mathcal{F}} \Expect \lVert R \xi - \mathbf{U} f \rVert^2$,
as shown in \cite{pytyev_ivs},
is minimal for $R$ that is equal to the linear unbiased reduction operator
\begin{equation}
\label{eqn:reduction-operator}
R_* \bydef \mathbf{U} (\mathbf{A}^* \mathbf{\Sigma}_{\nu}^{-1} \mathbf{A})^- \mathbf{A}^* \mathbf{\Sigma}_{\nu}^{-1},
\end{equation}
where ${}^-$ denotes pseudoinverse.
$h(R_*, \mathbf{U}) = \tr \mathbf{U} (\mathbf{A}^* \mathbf{\Sigma}_{\nu}^{-1} \mathbf{A})^{-1} \mathbf{U}^*$,
and the covariance matrix of the linear reduction estimate $R_* \xi$
is
\begin{equation}
\label{eqn:reduction-cov-op}
\mathbf{\Sigma}_{R_* \xi} = \mathbf{U} (\mathbf{A}^* \mathbf{\Sigma}_{\nu}^{-1} \mathbf{A})^{-1} \mathbf{U}^*.
\end{equation}

Estimation is possible (MSE is finite)
if the condition $\mathbf{U} (I - \mathbf{A}^- \mathbf{A}) = 0$ holds,
where, as noted above, $\mathbf{A}$ characterizes the \emph{real} measuring device,
while $\mathbf{U}$ characterizes an \emph{ideal} one
with the point spread function required by the researcher,
and, therefore, \emph{any desired resolution},
if this condition if fulfilled.
Note that, unlike fluorescence-based superresolution techniques,
see e.\,g. \cite{solomon_et_al_2018},
the proposed technique does not require attaching fluorescent molecules to the object.
However, as a rule,
the better the desired resolution of the ideal measuring device
compared to the resolution of the real one,
the larger MSE of the obtained estimate.
By choosing $\mathbf{U}$ one can select an acceptable (to him) compromise
between obtained resolution and noise magnitude.
In the case under consideration, as seen from \eqref{eqn:a-op-form},
diagonal elements of each block
(which, up to a nonzero factor, are equal to the factor before $|T(-\vr_j)|^2$ in the expression for correlation function $G_{j}$)
are nonzero.
Therefore, each block $\mathbf{A}_j$ is non-degenerate,
so for non-degenerate $\mathbf{B}_j$ the reduction error takes only finite values.
For a different multiplexing method and thus, different form of the matrix $\mathbf{A}$
this is generally not so.

The measurement reduction technique for the case when
it is known that the value $u$ of the feature of interest
is an arbitrary element not of the entire $\mathcal{U}$
but of its convex closed subset $\mathcal{U}_{\prior}$
was considered in \cite{reduction_vmu, lomo_readings}.
The estimate refinement which takes advantage of this information
is determined by solving the equation
\begin{equation}
\label{eqn:reduction-estimate-in-set}
\hat{u} = \Pi_{\mathbf{\Sigma}_{R_* \xi}}\left(\tilde{R}_{\mathbf{\Sigma}_{R_* \xi}} \left( \xi^T, \hat{u}^T \right)^T\right)
\end{equation}
for $\hat{u}$,
where
$\tilde{R}_{\mathbf{\Sigma}_{R_* \xi}}$ is the measurement reduction operator
for a MT $\left(\mathbf{A}^T, \mathbf{U}^T\right)^T$
and noise with covariance matrix $\begin{pmatrix}\mathbf{\Sigma}_{\nu} & 0\\ 0 & \mathbf{\Sigma}_{R_* \xi}\end{pmatrix}$,
and the operator
\begin{equation}
\label{eqn:mahalanobis-projection}
\Pi_{\mathbf{\Sigma}_{R_* \xi}}(u) \bydef
\argmin\limits_{v \in \mathcal{U}_{\prior}} (v - u, {\mathbf{\Sigma}_{R_* \xi}}^{-1} (v - u))
\end{equation}
describes projection onto $\mathcal{U}_{\prior}$
by minimizing Mahalanobis distance
$\lVert \mathbf{\Sigma}_{R_* \xi}^{-1/2} \cdot \rVert$
that is related to covariance matrix
$\mathbf{\Sigma}_{R_* \xi}$ \eqref{eqn:reduction-cov-op}
of error of the linear reduction estimate $R_* \xi$.
Note that the version of reduction technique proposed in \cite{ghost_images_jetp} and in \cite{reduction_vmu}
for similar information
used minimization of the ``ordinary'' Euclidean distance
instead of Mahalanobis distance.
In \cite{lomo_readings},
the advantages of minimizing Mahalanobis distance instead of Euclidean distance
during projection are shown.
In that case the covariance matrix \eqref{eqn:reduction-cov-op}
of linear reduction estimate error
is an upper bound on the covariance matrix of the obtained estimate.

\subsection{Representation of the object information that is available to the researcher}
\label{sec:image-information}

It is obvious that a priori $|T(-\vr_j)|^2 \in [0, 1]$,
hence $f \in [0, 1]^{\dim \mathcal{F}}$,
$\mathbf{U} f \in [0, 1]^{\dim \mathcal{F}}$.

It is assumed that the transparency distribution of the object is not ``entirely'' arbitrary:
transparencies of neighboring pixels usually do not differ much,
so the image is sparse (many of its components are zero) in a given (a priori known) basis,
similarly to compressed sensing ghost imaging
\cite{imaging_small_n_photons, hi-res_gi_sparsity, gi_compressed_sensing_substr_const}.

The researcher also knows the matrix $\mathbf{A}$ \eqref{eqn:a-op-form}
that describes image acquisition conditions
and, up to the vector $f$, the matrix $\mathbf{\Sigma}_{\nu}$ \eqref{eqn:sigma-op-form}
that describes measurement errors.
Note that the worst case of $f$ is realized if all pixels are equally transparent.
For a different multiplexing method
one considers the worst case in the sense of reduction MSE
of the object
in step \ref{itm:first-estimate} of the algorithm below.

\subsection{Reduction algorithm}
\label{sec:reduction-algorithm}

The proposed algorithm of
multiplexed GI processing using measurement reduction technique
that is based on the indicated prior information
has the following form.
\begin{enumerate}
\item
\label{itm:first-estimate}
Calculation of linear unbiased reduction estimate $R_* \xi$ \eqref{eqn:reduction-operator}
based on the acquired GI,
assuming for calculation of covariance matrix \eqref{eqn:sigma-op-form}
that all pixels have the same brightness.

\item
Refinement of the estimate $R_* \xi$ using the information $\mathcal{U}_{\prior} = [0, 1]^{\dim \mathcal{F}}$
by the method \eqref{eqn:reduction-estimate-in-set}
by fixed-point iteration,
i.\,e. by consecutive application of the mapping \eqref{eqn:reduction-estimate-in-set}
with $\Pi_{\mathbf{\Sigma}_{R_* \xi}} (R_* \xi)$ as the initial approximation.
We denote the obtained estimate by $\hat{u}$.

\item
Application of the sparsity-inducing transformation $T$ to $\hat{u}$.
``Sparsity-inducing'' means that
the transformation is chosen by the researcher
so that, in his opinion,
the transform of the true transparency distribution of the object
is sparse.

\item
\label{itm:thresholding}
Calculation of maximal (in the worst case of $f$) variances $\sigma_{T \hat{u}}^2 = (\sigma_{(T \hat{u})_1}^2, \dots, \sigma_{(T \hat{u})_{\dim \mathcal{F}}}^2)$
of the components of $T \hat{u}$,
i.\,e. the diagonal matrix elements of $T \mathbf{\Sigma}_{R_* \xi} T^*$,
and calculation of $T \hat{u}_{\textnormal{thr}}$:
$(T \hat{u}_{\textnormal{thr}})_i \bydef 0$ if $|(T \hat{u})_i| < \lambda \sigma_{(T \hat{u})_i}$,
otherwise $(T \hat{u}_{\textnormal{thr}})_i \bydef (T \hat{u})_i$.

\item
Inverse transformation $T^{-1}$ of $T \hat{u}_{\textnormal{thr}}$
(if $T$ is a unitary transformation, then $T^{-1} = T^*$),
i.\,e. calculation of
$\hat{u}_{\textnormal{thr}} \bydef T^{-1} T \hat{u}_{\textnormal{thr}}$.

\item
Calculation of the projection $\Pi_{\mathbf{\Sigma}_{R_* \xi}} (\hat{u}_{\textnormal{thr}})$
that is considered to be the result of processing obtained ghost images.
\end{enumerate}

The value of $\lambda \geq 0$ is a parameter of the algorithm.
It reflects a compromise between noise suppression
(the larger the value of $\lambda$, the greater the noise suppression)
and distortion of images whose components are close to $0$.
Step \ref{itm:thresholding} can be considered as testing a statistical hypothesis, according to which $(T \mathbf{U} f)_i = 0$
(for the alternative hypothesis that $(T \mathbf{U} f)_i \neq 0$)
for all $i$.
In this paper to do that we employ in step \ref{itm:thresholding} a simple criterion
based on Chebyshev's inequality:
if $(T \mathbf{U} f)_i = 0$,
then $\Pr\left(|(T \hat{u})_i| \geq \lambda \sigma_{(T \hat{u})_i}\right) \leq \lambda^{-2}$.
Due to that one can suppose that
image distortion is insignificant for, at least, $\lambda \leq 1$,
as such distortion would be indistinguishable from the noise.
Step \ref{itm:thresholding} can be also interpreted as replacement of
the original matrix $\mathbf{U}$ with one whose kernel contains
the estimate components after the specified transform
that are affected by the noise the most.

In \cite{ghost_images_jetp}
the matrix $\mathbf{U}$ was chosen
to suppress the noise more,
even at the cost of potential image distortion
(e.\,g. worse resolution),
by discarding the most noisy components of the image.
Unlike \cite{mgi_icono_lat_2016, mgi_correlations_jrlr_2017, ghost_images_jetp},
here we consider components of the image
in a basis specified by the researcher
instead of the eigenbasis \cite[ch.~8]{pytyev_ivs} of the measurement interpretation model,
i.\,e. a basis determined by error properties.
In this article the basis is defined by the transformation
whose result for the true transparency distribution is sparse,
but the discarded components are determined, as in \cite{ghost_images_jetp},
by the measurement error.
Thus, to improve estimation quality not only information about the noise
is used, but also information about the object,
namely, the properties of the transparency distribution
(in the opinion of the researcher) and its features of interest.

\section{Computer modeling results}
\label{sec:computer-modelling}

The results of processing of obtained GI as described above
are shown in fig.~\ref{fig:two-slits} and \ref{fig:phys-msu}.
The detectors in reference arms are identical ones
that are three times as large as an element of the object image.
Therefore, image processing via measurement reduction
increases resolution in addition to noise suppression.
Modeling was carried out for the same parameters of the optical setup
as in \cite{ghost_images_jetp}:
beam wave numbers $k_1 = 6 \cdot 10^4~\textnormal{cm}^{-1}$, $k_3 = 1.7 \cdot 10^5~\textnormal{cm}^{-1}$,
crystal parameter $\beta = 10~\textnormal{cm}^{-1}$,
crystal parameter $\xi = \gamma / \beta = 0.4$.

\setlength{\subfigsize}{0.24\linewidth}

\begin{figure}
\centering
\begin{subfigure}[t]{\subfigsize}
\centering
\includegraphics[scale=1]{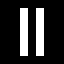}
\caption{Transparency distribution of the object}
\label{fig:two-slits-src}
\end{subfigure}
\begin{subfigure}[t]{2\subfigsize}
\centering
\includegraphics[scale=1]{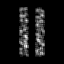}
\includegraphics[scale=1]{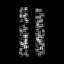}
\includegraphics[scale=1]{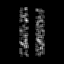}
\caption{GI}
\label{fig:two-slits-gi}
\end{subfigure}
\begin{subfigure}[t]{\subfigsize}
\centering
\includegraphics[scale=1]{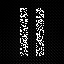}
\caption{Reduction result without sparsity information}
\label{fig:two-slits-red}
\end{subfigure}
\begin{subfigure}[t]{\subfigsize}
\centering
\includegraphics[scale=1]{{{two_slits-red-dct-1.25}}}
\caption{DCT, $\lambda = 1.25$}
\label{fig:two-slits-red-dct-1.25}
\end{subfigure}
\begin{subfigure}[t]{\subfigsize}
\centering
\includegraphics[scale=1]{{{two_slits-red-dct-2.0}}}
\caption{DCT, $\lambda = 2.0$}
\label{fig:two-slits-red-dct-2.0}
\end{subfigure}
\begin{subfigure}[t]{\subfigsize}
\centering
\includegraphics[scale=1]{{{two_slits-red-haar-1.0}}}
\caption{Haar transform, $\lambda = 1.0$}
\label{fig:two-slits-red-haar-1.0}
\end{subfigure}
\begin{subfigure}[t]{\subfigsize}
\centering
\includegraphics[scale=1]{{{two_slits-red-haar-2.0}}}
\caption{Haar transform, $\lambda = 2.0$}
\label{fig:two-slits-red-haar-2.0}
\end{subfigure}
\begin{subfigure}[t]{\subfigsize}
\centering
\includegraphics[scale=1]{{{two_slits-red-haar-3.0}}}
\caption{Haar transform, $\lambda = 3.0$}
\label{fig:two-slits-red-haar-3.0}
\end{subfigure}
\caption{GI processing by the developed algorithm:
(\subref{fig:two-slits-src}) is the object,
$64$x$64$ pixels,
that is illuminated by $1$ photon per pixel on average,
(\subref{fig:two-slits-gi}) are its acquired GI,
and (\subref{fig:two-slits-red}--\subref{fig:two-slits-red-haar-3.0}) are image reduction results:
(\subref{fig:two-slits-red})~is the result of reduction without sparsity information,
(\subref{fig:two-slits-red-dct-1.25}--\subref{fig:two-slits-red-haar-3.0})~are results of reduction using information about sparsity in
(\subref{fig:two-slits-red-dct-1.25}--\subref{fig:two-slits-red-dct-2.0})~discrete cosine transform (DCT),
(\subref{fig:two-slits-red-haar-1.0}--\subref{fig:two-slits-red-haar-3.0})~Haar transform
bases.
The parameter $\lambda \geq 0$ of the image processing algorithm
reflects a compromise between noise suppression
(the larger the value of $\lambda$, the greater the noise suppression)
and distortion of images whose components are close to $0$}
\label{fig:two-slits}
\end{figure}

\begin{figure}
\centering
\begin{subfigure}[t]{\subfigsize}
\centering
\includegraphics[scale=1]{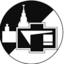}
\caption{Transparency distribution of the object}
\label{fig:phys-msu-src}
\end{subfigure}
\begin{subfigure}[t]{2\subfigsize}
\centering
\includegraphics[scale=1]{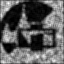}
\includegraphics[scale=1]{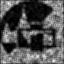}
\includegraphics[scale=1]{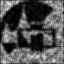}
\caption{GI}
\label{fig:phys-msu-gi}
\end{subfigure}
\begin{subfigure}[t]{\subfigsize}
\centering
\includegraphics[scale=1]{{{phys-msu-red-dct-1.0}}}
\caption{DCT, $\lambda = 1.0$}
\label{fig:phys-msu-red-dct-1.0}
\end{subfigure}
\begin{subfigure}[t]{\subfigsize}
\centering
\includegraphics[scale=1]{{{phys-msu-red-dct-1.5}}}
\caption{DCT, $\lambda = 1.5$}
\label{fig:phys-msu-red-dct-1.5}
\end{subfigure}
\begin{subfigure}[t]{\subfigsize}
\centering
\includegraphics[scale=1]{{{phys-msu-red-dct-2.0}}}
\caption{DCT, $\lambda = 2.0$}
\label{fig:phys-msu-red-dct-2.0}
\end{subfigure}
\begin{subfigure}[t]{\subfigsize}
\centering
\includegraphics[scale=1]{{{phys-msu-red-haar-1.0}}}
\caption{Haar transform, $\lambda = 1.0$}
\label{fig:phys-msu-red-haar-1.0}
\end{subfigure}
\begin{subfigure}[t]{\subfigsize}
\centering
\includegraphics[scale=1]{{{phys-msu-red-haar-1.5}}}
\caption{Haar transform, $\lambda = 1.5$}
\label{fig:phys-msu-red-haar-1.5}
\end{subfigure}
\begin{subfigure}[t]{\subfigsize}
\centering
\includegraphics[scale=1]{{{phys-msu-red-haar-2.0}}}
\caption{Haar transform, $\lambda = 2.0$}
\label{fig:phys-msu-red-haar-2.0}
\end{subfigure}
\caption{GI processing by the developed algorithm:
(\subref{fig:phys-msu-src}) is the object,
$64$x$64$ pixels,
that is illuminated by $10$ photons per pixel on average,
(\subref{fig:phys-msu-gi}) are its acquired GI,
and (\subref{fig:phys-msu-red-dct-1.0}--\subref{fig:phys-msu-red-haar-2.0})~are results of reduction using information about sparsity in
(\subref{fig:phys-msu-red-dct-1.0}--\subref{fig:phys-msu-red-dct-2.0})~discrete cosine transform (DCT),
(\subref{fig:phys-msu-red-haar-1.0}--\subref{fig:phys-msu-red-haar-2.0})~Haar transform
bases}
\label{fig:phys-msu}
\end{figure}

One can see that additional information about sparsity
allows to suppress noise more
but its impact on obtained resolution is weak.
As expected, for $0 \leq \lambda \leq 1$
the distortion is undistinguishable from the noise.
Further increase of $\lambda$ leads to better noise suppression
(cf., e.\,g., fig.~\ref{fig:two-slits-red} and \ref{fig:two-slits-red-dct-1.25}, \ref{fig:phys-msu-red-dct-1.0} and \ref{fig:phys-msu-red-dct-1.5}),
but also leads to more severe distortions
caused by discarding ``significant'' image components as well
(cf., e.\,g., fig.~\ref{fig:two-slits-red-dct-2.0} and \ref{fig:phys-msu-red-dct-2.0}).
For large $\lambda$,
their influence outweighs
the improvement of image quality due to noise suppression,
as small-scale image details are suppressed as well.
Therefore, the optimal value of $\lambda$ depends on one's intentions:
one should choose the maximal value of $\lambda$
that preserves the details of interest.
To do that, one can model acquisition of a test image
that contains the required details
and choose the largest value of $\lambda$
that preserves them,
or specify the value of $\lambda$
after comparing reduction results for different $\lambda$.
In the case of an object with sharp transparency changes (fig.~\ref{fig:two-slits})
the additional information allowed to suppress false signal
where the object is opaque,
but only for Haar transform
(discrete cosine transform (DCT) causes increased false signal in that region).

The transform whose result for the transparency distribution of the object
is sparse
that is usually employed in ghost image processing by the means of compressed sensing
is DCT
\cite{gi_compressed_sensing_substr_const,imaging_small_n_photons,hi-res_gi_sparsity}.
In \cite{sparsity_property_influence},
several transforms (identity transform, discrete wavelet transform and DCT)
were reviewed and the advantages of DCT were shown.
However, it seems that Haar transform may be preferable
in the case of a transparency distribution
that contains areas of weakly changing transparency with sharp borders
if these areas are large compared to the resolution of the ideal measuring transducer
and the location of the borders is important to the researcher.
This assumption is verified by fig.~\ref{fig:two-slits-red-haar-1.0}--\subref{fig:two-slits-red-haar-2.0},
where one can see that Haar transform in this case, as opposed to fig.~\ref{fig:phys-msu},
allows larger $\lambda$ values without causing significant distortions,
cf., e.\,g., fig.~\ref{fig:two-slits-red-dct-2.0} and \subref{fig:two-slits-red-haar-2.0},
where the usage of DCT causes blurring of transversal slit borders
for the same value of $\lambda$.

\begin{figure}
\centering
\begin{subfigure}[t]{\subfigsize}
\centering
\includegraphics[scale=1]{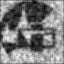}
\caption{Ordinary image}
\label{fig:phys-msu-extra-o-10}
\end{subfigure}
\begin{subfigure}[t]{\subfigsize}
\centering
\includegraphics[scale=1]{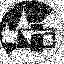}
\caption{Result of reduction of the ordinary image without sparsity information}
\label{fig:phys-msu-extra-o-10-red}
\end{subfigure}
\begin{subfigure}[t]{\subfigsize}
\centering
\includegraphics[scale=1]{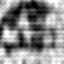}
\caption{DCT, $\lambda = 1.25$}
\label{fig:phys-msu-extra-o-10-dct}
\end{subfigure}
\begin{subfigure}[t]{2\subfigsize}
\centering
\includegraphics[scale=1]{{{phys-msu-extra-10-0.1-gi2}}}
\includegraphics[scale=1]{{{phys-msu-extra-10-0.1-gi3}}}
\includegraphics[scale=1]{{{phys-msu-extra-10-0.1-gi4}}}
\caption{GI + $1$ noise photons}
\label{fig:phys-msu-extra-gi-10-0.1}
\end{subfigure}
\begin{subfigure}[t]{\subfigsize}
\centering
\includegraphics[scale=1]{{{phys-msu-extra-10-0.1-red-dct}}}
\caption{DCT, $\lambda = 1.25$}
\label{fig:phys-msu-extra-red-dct-10-0.1}
\end{subfigure}
\begin{subfigure}[t]{2\subfigsize}
\centering
\includegraphics[scale=1]{{{phys-msu-extra-10-0.5-gi2}}}
\includegraphics[scale=1]{{{phys-msu-extra-10-0.5-gi3}}}
\includegraphics[scale=1]{{{phys-msu-extra-10-0.5-gi4}}}
\caption{GI + $5$ noise photons}
\label{fig:phys-msu-extra-gi-10-0.5}
\end{subfigure}
\begin{subfigure}[t]{\subfigsize}
\centering
\includegraphics[scale=1]{{{phys-msu-extra-10-0.5-red-dct}}}
\caption{DCT, $\lambda = 1.25$}
\label{fig:phys-msu-extra-red-dct-10-0.5}
\end{subfigure}
\caption{Ordinary and ghost image processing by the developed algorithm.
(\subref{fig:phys-msu-extra-o-10})~is the ordinary image of
the object from fig.~\ref{fig:phys-msu-src}
obtained by illuminating it by $10$ photons per pixel on average
and $10$ noise photons,
and (\subref{fig:phys-msu-extra-o-10-red}) and (\subref{fig:phys-msu-extra-o-10-dct})
are the results of its reduction
(\subref{fig:phys-msu-extra-o-10-dct})~without sparsity information
and (\subref{fig:phys-msu-extra-o-10-dct})~with information about sparsity in DCT base.
(\subref{fig:phys-msu-extra-gi-10-0.1}, \subref{fig:phys-msu-extra-gi-10-0.5}) are GI
impacted by $1$ and $5$ noise photons, respectively.
(\subref{fig:phys-msu-extra-red-dct-10-0.1}, \subref{fig:phys-msu-extra-red-dct-10-0.5})
are the results of their reduction with information about sparsity in DCT base}
\label{fig:gi-vs-ordinary-10}
\end{figure}

In fig.~\ref{fig:gi-vs-ordinary-10}
GI are compared with ordinary images
if noise photons, which do not carry information about the object,
but increase noise,
are present.
Due to employing correlations to acquire GI,
noise photons usually do not affect detected images,
as this requires simultaneous detection of a noise photon by one detector
and another photon by a different detector.
Nevertheless, due to finite coincidence windows
and finite widths of the light filters before the detectors
the noise photons do increase the measurement errors.
One can see that due to suppression of most noise photons
the quality of the reconstructed image
is better than the quality of the image reconstructed using the ordinary image
for the same number of noise photons.
Moreover, when taking advantage of sparsity information
ghost imaging allows to exploit larger $\lambda$ values
and thus, to suppress the noise more
(cf., e.\,g., fig.~\ref{fig:phys-msu-extra-o-10-dct}
and \ref{fig:phys-msu-extra-red-dct-10-0.5}).
In this case multiplexing provides the means for further noise suppression
if noise photons in different arms are detected independently.

Therefore,
formalization the researcher's information about
sparsity of the object transparency distribution
by Haar transform
is preferable if it has areas of weakly changing transparency with sharp borders
that are large compared to the resolution of the ideal measuring transducer
and the location of the borders is important to the researcher.
DCT is preferable if the transparency distribution
has small transparency changes that have to be present in the estimate,
e.\,g. biological objects without high-contrast borders.
The values of $\lambda \sim 1 \div 1.5$ are optimal if small-scale details
are present and are of interest.
Otherwise, larger $\lambda$ values are advisable.

\section*{Conclusion}

The actual problem of increasing noise immunity is exacerbated
by photon transmission and detecton in photocounting mode
due to higher information content of each photon or its absense.
Multiplexing of ghost images allows to reduce the noise level,
since it increases the amount of transmitted information,
enabling improvement the quality of processing of acquired data.
In this case the additional information available to the researcher
about the measurement process and about the object
allows further noise suppression under the same detection conditions.
Alternatively, one can make the detection conditions worse
(e.\,g. to reduce the number of photons)
while preserving the same estimation quality.
The additional information about the measuring process in this work
is the correlation functions of multiplexed ghost images.
The additional information about the object
is the information that the object transparency distribution is not arbitrary,
namely, transparencies of neighboring pixels, as a rule, differ only slightly.
This information is formalized as sparsity of the result of a given transform
(e.\,g. DCT)
of the transparency distribution,
similar to compressed sensing.

In compressed sensing, as a rule, the measurement error is modeled
as an arbitrary vector with bounded norm.
Instead, in the proposed method it is modeled as a random vector,
and selection of the estimate components which are considered to be zero
is based on the statistical properties of the estimate components,
namely, their variances.
The use of covariances of the estimate components in addition to their variances
is a subject of further research.

We consider that computer modeling based on the developed algorithm
showed high efficiency of the developed reduction technique of ghost image processing
in the sense of improvement of both their quality
and their noise immunity.
It is of interest to apply this technique in the field of quantum image processing
for parametric amplification of images and frequency conversion.

The authors are grateful for help to T.\,Yu.\,Lisovskaya.
This work was supported by RFBR grant 18-01-00598-A.

This is a pre-print of an article published in Quantum Information Processing. The final authenticated version is available online at: \url{https://doi.org/10.1007/s11128-019-2193-x}

\bibliographystyle{unsrt}
\bibliography{gi2018bibliography}
\end{document}